\documentclass[12pt]{article}
\usepackage{epsfig}

\newcommand{\gtrsim}
{\mathrel{\raisebox{-2.8pt}{\mbox{$\stackrel{\textstyle >}{\sim}$}}}}
\makeatletter
\long\def\@makecaption#1#2{{\small
\advance\leftskip1cm
\advance\rightskip1cm
\vskip\abovecaptionskip
\sbox\@tempboxa{#1: #2}%
\ifdim \wd\@tempboxa >\hsize
 #1: #2\par
\else
\global \@minipagefalse
\hb@xt@\hsize{\hfil\box\@tempboxa\hfil}%
\fi
\vskip\belowcaptionskip}}
\makeatother

\newcommand{\abs}[1]{\left|#1\right|}

\newcommand{\rbk}[1]{\left(#1\right)}

\newcommand{\bkt}[1]{\left\langle#1\right\rangle}

\newcommand{\rhom}{\rho_{\rm m}}
\newcommand{\rhop}{\rho_{\rm p}}
\newcommand{\rhoi}{\rho_{\rm i}}
\newcommand{\pppp}{{\bf p}^{(2)}}
\newcommand{\rppp}{{\bf r}^{(2)}}
\newcommand{\gammappp}{\gamma^{(2)}}
\newcommand{\tip}{\tilde{\bf p}}
\newcommand{\tir}{\tilde{\bf r}}
\newcommand{\tigamma}{\tilde{\gamma}}
\newcommand{\tiGamma}{\tilde{\Gamma}}
\newcommand{\tix}{\tilde{x}}
\newcommand{\hatp}{\hat{\bf p}}
\newcommand{\hatr}{\hat{\bf r}}
\newcommand{\hatgamma}{\hat{\gamma}}
\newcommand{\hatGamma}{\hat{\Gamma}}
\newcommand{\bp}{{\bf p}}
\newcommand{\br}{{\bf r}}

\newcommand{\pth}{p_{\rm th}}
\newcommand{\eR}{e_{\rm R}}
\newcommand{\DE}{\Delta E}
\newcommand{\DtE}{\Delta\tilde{E}}
\newcommand{\eqref}[1]{(\ref{e:#1})}

\begin{document}
\noindent
{\Large\bf The coefficient of restitution does not exceed unity}

\setcounter{footnote}{1}
\footnotetext{This version will appear in Journal of Statistical Physics.}
\bigskip\noindent
Hal Tasaki\footnote{
Department of Physics, Gakushuin University,
Mejiro, Toshima-ku, Tokyo 171-8588, Japan,
{\tt hal.tasaki\makeatletter @\makeatother gakushuin.ac.jp}
}

\begin{abstract}
We study a classical mechanical problem in which a macroscopic  ball is reflected by a non-deformable wall.
The ball is modeled as a collection of classical particles bound together by an arbitrary potential, and its internal degrees of freedom are initially set to be in thermal equilibrium.
The wall is represented by an arbitrary potential which is translation invariant in two directions.
We then prove that the final normal momentum can exceed the initial normal momentum  at most  by $O(\sqrt{mkT})$, where $m$ is the total mass of the ball, $k$ the Boltzmann constant, and $T$ the temperature.
This implies the well-known statement in the title in the macroscopic limit where $O(\sqrt{mkT})$ is negligible.
Our result may be interpreted as a rigorous demonstration of the second law of thermodynamics in a system where a macroscopic dynamics and microscopic degrees of freedom are intrinsically coupled with each other.
\end{abstract}

\section{Introduction}
Impact between macroscopic bodies is a common phenomenon in which the complicated dynamics of microscopic degrees of freedom plays an essential role in determining the macroscopic motions of the bodies.
A standard quantity characterizing impact is the coefficient of restitution $\eR$, which is the ratio between the normal (relative) velocities before and after the impact.
Although some elementary physics textbooks refer to the coefficient of restitution $\eR$ as a material constant, careful studies have revealed that $\eR$ depends sensitively on various conditions \cite{eR1,eR2}.
It has even been  pointed out that the generally believed rule that $\eR$ cannot exceed unity may be violated in oblique impacts between a ball and a wall, provided that the wall is locally deformable \cite{egt1a,egt1b}.

\begin{figure}
\begin{center}
\epsfig{file=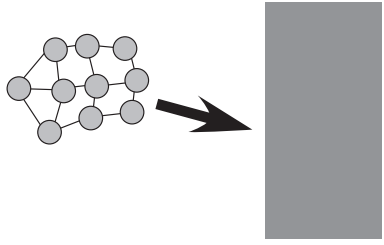,width=5cm}
\end{center}
\caption{
A ``ball'' consisting of $N$ classical particle is reflected by a non-deformable wall.
}
\label{f:system}
\end{figure}

Given such recent progress, it seems necessary to develop theories of impact from fundamental and statistical mechanical points of view.
Here {\em we prove a basic theorem which establishes the above mentioned rule $\eR\le1$ (in the macroscopic limit) based on microscopic dynamics in a general and physically natural setting}\/.
More precisely we model a non-deformable wall by an arbitrary fixed potential which is translation invariant in two directions, and a ball as a collection of $N$ classical particles which are bound together by an arbitrary potential.
We assume that the internal degrees of freedom of the ball are initially in their thermal equilibrium, and give the center of mass of the ball a macroscopic momentum.
See Fig.~\ref{f:system}.
Then we prove rigorous inequalities about the final momentum, which essentially implies $\eR\le1$ for a macroscopic ball.
Our bound contains a small correction (which is negligible for macroscopic bodies)
which shows that even very small bodies can violate the $\eR\le1$  rule only
by a universal factor determined by the thermal momentum.

The rule $\eR\le1$  itself is by no means surprising.
But as far as we know such a fundamental and well known relation has not been established rigorously before.
We believe that our explicit treatment of  interplay between microscopic and macroscopic degrees of freedom in a dynamical system is unique and new.

Of course the rule $\eR\le1$ is closely related to the second law of thermodynamics.
In the present problem ({\em not}\/ in those treated in \cite{egt1a,egt1b}), that $\eR\le1$ essentially means that the internal thermal energy of the ball cannot be converted into macroscopic kinetic energy.
This can be regarded as a manifestation of the second law in the form of Kelvin's law.
We stress, however, that although there have been some rigorous derivations of the second law of thermodynamics based on microscopic mechanics \cite{second1,second2,second3,second4,second5,second6,second7}, none of them apply to the present situation of a ball reflected by a wall.
The reason for the inapplicability is essential rather than technical.
In the existing proofs of the second law, macroscopic degrees of freedom are assumed to be completely controlled by an outside agent.
The agent performs an  operation (represented by a time-dependent Hamiltonian) according to a pre-fixed protocol, without allowing any feedback from the microscopic degrees of freedom\footnote{
In thermodynamic settings, it is implicitly assumed that the feedback from microscopic degrees of freedom becomes negligible if the system and the means of operations become macroscopic.
It is hard to judge how realistic such an idealization is.
}.
In the present problem of impact, on the other hand, the macroscopic motion of the ball is determined only as a solution to the equation of motion involving all the degrees of freedom.
It is therefore mandatory here to take into account feedback to the macroscopic motion from the microscopic motion.
We accomplish this by making a full use of the symmetry of the model.

Thus, from the view points of statistical mechanics and thermodynamics, our result $\eR\le1$ can be regarded as {\em a step in establishing the second law by properly taking into account feedback from microscopic degrees of freedom to macroscopic motion}\/.
Such a treatment is not only realistic but also sheds better light on the interplay between the macroscopic and the microscopic scales.
We also note that in small physical systems (such as molecular motors) which operate under relatively large thermal fluctuation, feedback from internal degrees of freedom to overall motion plays an essential role.
Recently Maes and the present author extended the present work, and proved similar (but weaker) results in a much more general setting \cite{MaesTasaki}.

We note that the second law has been extended to nonequilibrium states in \cite{GL,GGL} by examining the Boltzmann entropy.
Although these works do not treat concrete mechanical setup as our, they deal with very similar problem as the present work, namely, to demonstrate the second law by starting from a nonequilibrium state and following the deterministic dynamics.
The constancy of the phase space volume plays an essential roles both in \cite{GL,GGL} and the present work (as well as in the Boltzmann's original work).

The reader might ask what our result imply about the thermalization process which should take place within the ball during and after the impact.
Interestingly enough our analysis does not deal directly with possible thermalization, and still rigorously leads us to the desired rule $\eR\le1$.
This is indeed physically natural since a thermalization process is not universal while the rule $\eR\le1$  is.
Although it is likely that usually a sufficient thermalization takes place during the impact, this may not be the case when the impact occurs very quickly.
Nevertheless the rule $\eR\le1$  is always valid, and should be provable.
It is of course challenging to study thermalization, but it requires a much more detailed and concrete analysis.

\section{Definitions and main results}
We consider a classical ``ball'' 
which consists of $N$ point-like particles
whose canonical  momenta and coordinates are
denoted as $\bp_{i}$ and $\br_{i}$,
respectively,
for $i=1,2,\ldots,N$.
We use the Cartesian coordinates when expressing the momenta and coordinates in their components.
The Hamiltonian of the 
ball is 
\begin{equation}
H=
\sum_{i=1}^N\frac{|\bp_{i}|^2}{2\mu_{i}}
+
\sum_{i>j}U_{i,j}(\br_{i}-\br_{j})
+
\sum_{i=1}^NV_{i}(\br_{i}),
\label{e:H1}
\end{equation}
where \( \mu_{i} \) is the mass of the \( i \)-th
particle, \( U_{i,j}(\cdot) \) is the potential of
the force between the particles $i$ and $j$,
and \( V_{i}(\cdot) \) is the potential
of the force that the wall exerts on the
\( i \)-th particle.
We assume that each $V_i(\cdot)$ is invariant under any translation in the
y or z directions.
Other assumptions about the potentials of the wall will be stated later.
As for the interaction, 
we assume that  $U_{i,j}(\br_i-\br_j)\ge U_0$ with a constant $U_0$, and $U_{i,j}(\br_i-\br_j)=\infty$ when $|\br_i-\br_j|\ge R_0$ with a constant  $R_0>0$.
The first assumption is introduced to guarantee that the canonical distribution \eqref{e:rig} is well-defined.
It may be replaced by a weaker condition if necessary.
The second assumption guarantees that the ball cannot be larger than $R_0$.
This constraint makes the proof less technical.
Otherwise we  have no  requirements for \( U_{i,j}(\cdot) \).

We stress that the ball need not be a uniform one.
One can design suitable $U_{i,j}(\cdot)$ (and constraints of the configuration if necessary) to treat, for example, an elastic sphere filled with a gas.
Moreover the ball may have an arbitrary shape (determined by the potential $U_{i,j}$), so it need not be a ``ball.''

As usual we introduce the momentum\footnote{
We write $A:=B$ when we want to emphasize that $B$ defines $A$.
}
\begin{equation}
\bp=(p_{x},p_{y},p_{z}):=\sum_{i=1}^N\bp_i,
\label{e:bp}
\end{equation}
and the coordinate 
\begin{equation}
\br=(x,y,z):=\frac{1}{m}\sum_{i=1}^N\mu_i\,\br_i,
\label{e:br}
\end{equation}
of the center of mass,
where \( m=\sum_{i}\mu_{i} \)
is the total mass of the ball.
We also denote by \( \gamma \) the collection of 
canonical coordinates for the remaining \( 3(N-1) \) 
internal 
degrees of freedom.
We denote these variables collectively as $\Gamma=(\bp,\br,\gamma)$ and
denote by $d\Gamma=d\bp\,d\br\,d\gamma$ the Lebesgue measure on the
whole $6N$-dimensional phase space.

By using these new variables, we write the potential from the wall as
\begin{equation}
V(\br,\gamma)=\sum_{i=1}^NV_i(\br_i).
\label{e:Vrg}
\end{equation}
Then the Hamiltonian (\ref{e:H1})
is rewritten as
\begin{equation}
    H=\frac{|\bp|^2}{2m}
    +h(\gamma)+V(\br,\gamma),
    \label{e:E}
\end{equation}
where \( h(\gamma) \) is the Hamiltonian for the
internal degrees of freedom.
Let us list our assumptions about the potential $V(\br,\gamma)$.

\noindent
i) $V(x,y,z,\gamma)$ is independent of $y$ and $z$.  

\noindent
ii) $V(x,y,z,\gamma)=0$ if $x\le0$ for any $\gamma$ such that $h(\gamma)<\infty$.

\noindent
iii) For any $\gamma$ such that $h(\gamma)<\infty$, one has $\frac{\partial}{\partial x}V(x,y,z,\gamma)\ge 0$ for any $x\ge0$ and $\frac{\partial}{\partial x}V(x,y,z,\gamma)\ge f_0$ for $x\ge r_0$.
Here $f_0>0$ and $r_0>0$ are constants.

The translation invariance i) is a consequence of the already stated translation invariance of $V_i(\br_i)$.
The property ii) means that the effect of the wall takes place only for
sufficiently large $x$.
As a consequence, the internal degrees of freedom of the ball and the motion of its center of mass are decoupled if $x<0$.
Then the center of mass moves with a constant velocity and the internal degrees of freedom $\gamma$ evolve separately according to the nontrivial internal dynamics determined by $h(\gamma)$.
The existence of such a ``decoupled motion'' plays an essential role in our proof.
The property iii) means that the wall exerts repulsive force which becomes strong enough if the ball goes deeper into the wall.
This guarantees that the ball is reflected by the wall.

\begin{figure}
\begin{center}
\epsfig{file=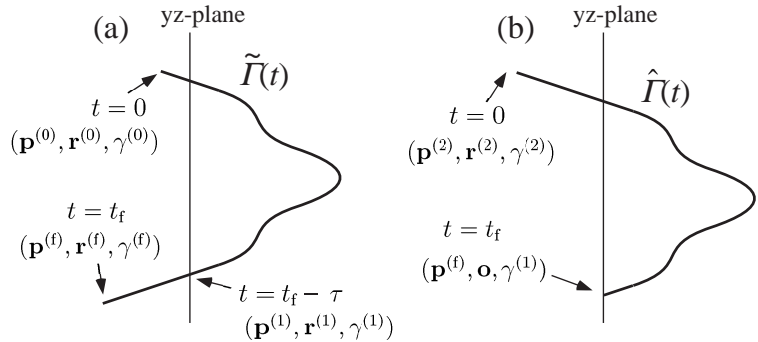,width=10cm}
\end{center}
\caption{
The solutions $\tiGamma(t)$ and $\hatGamma(t)$ are related by a shift of time and a translation.
The variables in the figure play an important role in the proof.
}
\label{f:variables}
\end{figure}

We denote by $\tiGamma(t)=(\tip(t),\tir(t),\tigamma(t))$ a 
solution of the equation of motion determined by $H$.
Of course this is a full fledged $N$-body problem, and one can never write down the solution $\tiGamma(t)$ explicitly.
Nevertheless we will be able to control some aspects of the motion rigorously.
We denote the initial state
as $\Gamma^{(0)}=(\bp^{(0)},\br^{(0)},\gamma^{(0)}):=\tiGamma(0)$.
See Fig.~\ref{f:variables}~(a).

Let us specify the initial condition.
Roughly speaking we take the initial state $\Gamma^{(0)}=(\bp^{(0)},\br^{(0)},\gamma^{(0)})$ 
where the momentum $\bp^{(0)}$ of the center of mass (almost) 
has a macroscopic value $\bp_0$, and the internal degrees of 
freedom of the ball are in equilibrium.
Therefore the ball initially has no macroscopic rotation.
More precisely the initial state $\Gamma^{(0)}$ is sampled
according to a measure 
$\rho(\Gamma)\,d\Gamma$,
where the weight $\rho(\Gamma)$ is written in a product form as
\begin{equation}
\rho(\Gamma)=\rhom(\bp)\,\rhop(\br)\,\rhoi(\gamma).
\label{e:rhoGamma}
\end{equation}
Here the probability $\rhop(\br)$ for the position is an arbitrary normalized density
(including delta functions) such that $\rhop(x,y,z)=0$ if $x>0$.
Thus the ball does not feel the wall at $t=0$.
As for the momentum we use for simplicity the normalized density given by 
\begin{equation}
\rhom(\bp)=
\cases{
3/\{4\pi(\Delta p)^3\}&if $|\bp-\bp_0|\le\Delta p$;\cr
0&otherwise,
}
\label{e:rm}
\end{equation}
where the initial macroscopic momentum is $\bp_0=(p_0,p_1,p_2)$ with $p_0>2\Delta p>0$ and arbitrary $p_1$, $p_2$.
Finally we assume that the internal degrees of freedom of the ball are in equilibrium,
 and sample them according to the canonical distribution
\begin{equation}
\rhoi(\gamma)=\frac{e^{-\beta\,h(\gamma)}}{Z(\beta)}
\label{e:rig}
\end{equation}
with the partition function
\begin{equation}
Z(\beta)=\int d\gamma\,e^{-\beta\,h(\gamma)},
\label{e:Zbeta}
\end{equation}
where $\beta=(kT)^{-1}$ is the inverse temperature.
Note that the canonical distribution is the physically correct choice (even for a very small ball) if the ball was in touch with a heat bath before the collision experiment\footnote{
To be precise one must assume that the interaction between the ball and the bath is very weak, and that one can detach the ball from the bath without disturbing the state of the ball.
}.
When the ball can be regarded as a ``healthy'' macroscopic body, the choice of the equilibrium ensemble should not matter because of the ``equivalence of ensemble.''
To demonstrate this, we treat in section~\ref{s:MC} the situation where the internal degrees of freedom are initially distributed according to the microcanonical ensemble.

Let $t_{\rm f}>0$ be a time such that $x(t)<0$ for any $t\ge t_{\rm f}$ with probability one (for the initial state as described above).
 In other words the ball has certainly been reflected by the wall and 
 is already in a ``decoupled motion'' at time $t_{\rm f}$.
 Note that such $t_{\rm f}$ always exists because of the assumed property iii) about the force from the wall\footnote{
We have assumed iii) only to ensure the existence of a finite $t_\mathrm{f}$.
Thus one can replace iii) with weaker conditions which still ensure the existence of $t_\mathrm{f}$.
}.
 We shall fix $t_{\rm f}$ throughout the paper, and denote the final state as $\Gamma^{(\mathrm{f})}=(\bp^{(\mathrm{f})},\br^{(\mathrm{f})},\gamma^{(\mathrm{f})}):=\tiGamma(t_{\rm f})$.
 See Fig.~\ref{f:variables}~(a).

 For any $p>0$, 
 let $P_\mathrm{f}(p)$ be the probability that the final momentum $\bp^{(\mathrm{f})}=(p_x^{(\mathrm{f})},p_y^{(\mathrm{f})},p_z^{(\mathrm{f})})$ satisfies
 $|p_x^{(\mathrm{f})}|\ge p$.
 We define the thermal momentum as $\pth=\sqrt{m/\beta}$,
 which is extremely small for a macroscopic ball.
 Then the main result of the present paper is the following.
 
 \bigskip
 \noindent
 {\bf Theorem:}
 Set the uncertainty in the initial momentum as $\Delta p =\pth$.
 Then one has
 \begin{equation}
P_\mathrm{f}(p)\le 12\frac{(\pth+p)^3}{p_0(\pth)^2}\exp[
-\frac{\beta}{2m}\{p^2-(p_0+\pth)^2\}],
\label{e:Pp0}
\end{equation}
and for the coefficient of restitution 
\begin{equation}
\eR:=\frac{\langle|p_x^{(\mathrm{f})}|\rangle}{p_0}
\le
1+2\frac{\pth}{p_0}+325\rbk{\frac{p_0}{\pth}}^2
\exp[-\frac{p_0}{\pth}].
\label{e:R0}
\end{equation}
\bigskip

The bound (\ref{e:R0})  shows that the coefficient of restitution $\eR$ does not exceed unity by much when $\pth\ll p_0$.
Note that the condition $\pth\ll p_0$ is always met in a macroscopic ball.
The bound (\ref{e:Pp0}) contains stronger information, and
shows that it is exponentially unlikely to have a final momentum $\bp^{(\mathrm{f})}$ with $|p^{(\mathrm{f})}_x|-p_0\gtrsim\pth$.

When the ball consists of a small number of particles and the thermal momentum $\pth$ is not negligible (compared with $p_0$), there can be some events\footnote{
One can easily construct such events with $|p_x^{(\mathrm{f})}|>p_0$, for example, in a model of two particles coupled by a harmonic potential.
} where the final momentum $|p_x^{(\mathrm{f})}|$ becomes strictly larger than $p_0$.
But our  bound (\ref{e:Pp0}) is valid even for such small balls, and establishes that {\em the final momentum $|p_x^{(\mathrm{f})}|$ may exceed $p_0$ only by a factor comparable to $\pth$}\/.
This means that when we say that a ball is macroscopic, we should speak about whether the initial momentum $p_0$  is large compared with the thermal momentum $\pth$, rather than whether the ball consists of many particles.

The requirement  $\Delta p=\pth$ in the theorem may sound natural from a practical point of view since it would be very hard to make $\Delta p$ smaller than $\pth$.
Theoretically speaking, we expect that the inequalities (\ref{e:Pp0}) and (\ref{e:R0}) (with slight improvements) are valid in the limit $\Delta p\to0$ as well, but cannot prove them for purely technical reasons.
The technical difficulty appears when we try to decouple macroscopic degrees of freedom (i.e., the $\bp^{(\mathrm{f})}$ integral) from microscopic degrees of freedom (i.e., the $\gamma^{(1)}$ integral) in (\ref{e:Pp3}).
It is desirable to find an improved proof which works in the $\Delta p\to0$ limit.

\section{The case with a fixed protocol}
Before describing the proof, which is rather involved, let us  
describe the corresponding argument which works in the case 
where the macroscopic motion is treated as a fixed protocol
as in the traditional proof of the second law \cite{second1,second2,second3,second4,second5,second6,second7}.
We hope that this short description will make clear both the essence of the proof and the difficulties encountered in the fully mechanical problem that we treat.

Suppose that the internal degrees of freedom are governed by a time-dependent Hamiltonian $h(t;\gamma)$ such that $h(0;\gamma)=h(t_\mathrm{f};\gamma)=h(\gamma)$.
One may imagine that an outside agent performs a fixed operation on the ball.
Again we denote by $\gamma^{(0)}$ and $\gamma^{(\mathrm{f})}$ the states at time $t=0$ and $t=t_\mathrm{f}$, respectively.
From the energy conservation law, the energy that the external agent gains during the operation is
\begin{equation}
\DE:=h(\gamma^{(0)})-h(\gamma^{(\mathrm{f})}).
\label{e:DE}
\end{equation}

Suppose that the initial state $\gamma^{(0)}$ is distributed according to $\rhoi(\gamma^{(0)})$ of \eqref{rig}.
For any $D\ge0$, let $\tilde{P}(D)$ be the probability that $\Delta E\ge D$.
Then from the definition, one has
\begin{equation}
\tilde{P}(D)
=\int_{\Delta E\ge D}d\gamma^{(0)}\,\frac{e^{-\beta\,h(\gamma^{(0)})}}{Z(\beta)}.
\label{e:idea1}
\end{equation}
Now the Liouville theorem implies $d\gamma^{(0)}=d\gamma^{(\mathrm{f})}$.
Then, from the energy conservation law \eqref{DE}, 
the probability (\ref{e:idea1}) can be bounded as
\begin{equation}
\tilde{P}(D)
=\int_{\Delta E\ge D}d\gamma^{(\mathrm{f})}\,
\frac{e^{-\beta\,h(\gamma^{(\mathrm{f})})-\beta \Delta E}}{Z(\beta)}
\le
\int d\gamma^{(\mathrm{f})}\,\frac{e^{-\beta\,h(\gamma^{(\mathrm{f})})-\beta D}}{Z(\beta)}
=e^{-\beta\,D}.
\label{e:idea2}
\end{equation}
The final factor decays rapidly for $D\gtrsim1/\beta$, and implies that the energy gain $\DE$ cannot be larger than $O(kT)$.
In the macroscopic limit (where $O(kT)$ is negligible), this implies  Kelvin's version of second law, i.e., $\DE\le0$.

The above argument, which makes use only of the energy conservation law and the Liouville theorem, contains the basic strategy of our proof of the main theorem.
But it is obvious that the same argument cannot be applied  to our fully mechanical problem of the ball reflected by the wall.
The main difficulty comes from the fact that the macroscopic motion itself becomes a dynamical degree of freedom, and the Liouville theorem does not imply  $d\gamma^{(0)}=d\gamma^{(\mathrm{f})}$, but  implies $d\Gamma^{(0)}=d\Gamma^{(\mathrm{f})}$.
To overcome the difficulty, we shall make full use of the symmetry of the model and the property of the space of the momentum of the center of mass.

\section{Proof}
Note that we can set $p_1=p_2=0$ by making a suitable Galilei transformation.
Then the initial momentum is $\bp_0=(p_0,0,0)$ plus a small fluctuation.

Since the initial measure $\rho(\Gamma)$ is normalized, the probability in question is written as
\begin{equation}
P_\mathrm{f}(p)=\int_{|p_x^{(\mathrm{f})}(\Gamma^{(0)})|\ge p}d\Gamma^{(0)}\,\rho(\Gamma^{(0)}),
\label{e:Pp01}
\end{equation}
where $\Gamma^{(\mathrm{f})}(\Gamma^{(0)})=(\bp^{(\mathrm{f})}(\Gamma^{(0)}),\br^{(\mathrm{f})}(\Gamma^{(0)}),\gamma^{(\mathrm{f})}(\Gamma^{(0)}))$ is the final state represented as a function of the initial state $\Gamma^{(0)}$.
Since $\Gamma^{(0)}$ and $\Gamma^{(\mathrm{f})}$ are in one-to-one correspondence and the Liouville theorem implies $d\Gamma^{(0)}=d\Gamma^{(\mathrm{f})}$, one can rewrite (\ref{e:Pp01}) as
\begin{equation}
P_\mathrm{f}(p)=\int_{|p_x^{(\mathrm{f})}|\ge p}d\Gamma^{(\mathrm{f})}\,\rho(\Gamma^{(0)}(\Gamma^{(\mathrm{f})})),
\label{e:Pp1}
\end{equation}
where $\Gamma^{(0)}(\Gamma^{(\mathrm{f})})$ is the initial state represented as a function of the final state $\Gamma^{(\mathrm{f})}$.
Since $\Gamma^{(0)}=(\bp^{(0)},\br^{(0)},\gamma^{(0)})$ depends on $\Gamma^{(\mathrm{f})}=(\bp^{(\mathrm{f})},\br^{(\mathrm{f})},\gamma^{(\mathrm{f})})$ in a nontrivial manner, it is not easy to evaluate the integral (\ref{e:Pp1}) as it is.
To ``disentangle'' the dependence as much as possible, we make a change of variables.

Recall that $\tiGamma(t)=(\tip(t),\tir(t),\tigamma(t))$ is a solution of the equation of motion with $\tiGamma(0)=\Gamma^{(0)}$ and $\tiGamma(t_{\rm f})=\Gamma^{(\mathrm{f})}$.
Since $x^{(\mathrm{f})}=\tix(t_{\rm f})<0$, there is a $\tau>0$ such that $\tix(t_{\rm f}-\tau)=0$, where we wrote $\tir(t)=(\tix(t),\tilde{y}(t),\tilde{z}(t))$.
Indeed there are two values of $\tau$ satisfying $\tix(t_{\rm f}-\tau)=0$, and we take the smaller one.
In other words, the center of mass passes through the yz-plane (defined by $x=0$) at $t=t_{\rm f}-\tau$ {\em after}\/ the reflection from the wall.
Since the motion ``decouples'' for $t\ge t_{\rm f}-\tau$, the center of mass has a constant velocity $\bp^{(\mathrm{f})}/m$, and we have $\tau=mx^{(\mathrm{f})}/p_x^{(\mathrm{f})}$.
Let us write $\Gamma^{(1)}=(\bp^{(1)},\br^{(1)},\gamma^{(1)}):=\tiGamma(t_{\rm f}-\tau)$.
We then find $\bp^{(1)}=\bp^{(\mathrm{f})}$, $\br^{(1)}=\br^{(\mathrm{f})}-(\bp^{(\mathrm{f})}/m)\tau$, and (by definition) $x^{(1)}=0$.
See Fig.~\ref{f:variables}~(a).

Let us consider another solution $\hatGamma(t)=(\hatp(t),\hatr(t),\hatgamma(t))$ (which satisfies the same equation of motion determined by $H$) with the ``final condition'' $\hatGamma(t_{\rm f})=(\bp^{(\mathrm{f})},{\bf o},\gamma^{(1)})$ where ${\bf o}=(0,0,0)$.
We then trace back the time evolution and write $\Gamma^{(2)}=(\pppp,\rppp,\gammappp):=\hatGamma(0)$.
See Fig.~\ref{f:variables}~(b).
We can naturally regard $\pppp$, $\rppp$, and $\gammappp$ as functions only of $\bp^{(\mathrm{f})}$ and $\gamma^{(1)}$.

Note that the two solutions $\tiGamma(t)$ and $\hatGamma(t)$ share the momentum $\bp^{(\mathrm{f})}=\bp^{(1)}$ and the internal coordinate $\gamma^{(1)}$ at the moment when the center of mass passes through the yz-plane after the impact.
Since the whole system is translation invariant in the y and the z directions, this means that the two solutions are related by a simple shift of time and a spatial translation as $\tip(t)=\hatp(t+\tau)$, $\tir(t)=\hatr(t+\tau)+\br^{(1)}$, and $\tigamma(t)=\hatgamma(t+\tau)$.
Noting that the motion ``decouples'' before the center of mass passes the yz-plane, these relations imply 
\begin{equation}
\bp^{(0)}=\tip(0)=\hatp(\tau)=\pppp,
\label{e:ppp}
\end{equation}
 and 
\begin{equation}
\gamma^{(0)}=\tigamma(0)=\hatgamma(\tau)=F_\tau(\gammappp),
\label{e:ggg}
\end{equation}
 where $F_\tau(\cdot)$ represents the time evolution generated by $h(\gamma)$ alone.
As for the positions $\br^{(0)}$ and $\rppp$, we use  $\hatr(\tau)=\hatr(0)+(\bp/m)\tau$ to get 
\begin{eqnarray}
\br^{(0)}&=&\tir(0)=\hatr(\tau)+\br^{(1)}
\nonumber\\
&=&\hatr(0)+\frac{\bp^{(0)}}{m}\tau+\br^{(1)}
\nonumber\\
&=&\rppp+\frac{\bp^{(0)}}{m}\tau+\br^{(\mathrm{f})}-\frac{\bp^{(\mathrm{f})}}{m}\tau
\nonumber\\
&=&
\rppp+\br^{(\mathrm{f})}
+\frac{\tau}{m}(\bp^{(0)}-\bp^{(\mathrm{f})}).
\label{e:qtq0}
\end{eqnarray}
We then substitute $\tau/m=x^{(\mathrm{f})}/p_x^{(\mathrm{f})}$ to get
\begin{equation}
\br^{(0)}=
\rppp+(0,y^{(\mathrm{f})},z^{(\mathrm{f})})+\frac{x^{(\mathrm{f})}}{p_x^{(\mathrm{f})}}(p_x^{(0)},p_y^{(0)}-p_y^{(\mathrm{f})},p_z^{(0)}-p_z^{(\mathrm{f})}).
\label{e:qtq}
\end{equation}
Suppose that $\bp^{(\mathrm{f})}$ and $\gamma^{(1)}$ are fixed, and consequently, $\bp^{(2)}$ and $\br^{(2)}$ are also fixed.
Then \eqref{qtq0} implies that the coordinate variables $\br^{(0)}$ and $\br^{(\mathrm{f})}$ are related with each other by a simple linear transformation.
This is the most important observation in the proof.

Since $\gamma^{(\mathrm{f})}=F_\tau(\gamma^{(1)})$, the Liouville theorem for the internal degrees of freedom implies $d\gamma^{(\mathrm{f})}=d\gamma^{(1)}$.
We also have $\rhoi(\gamma^{(0)})=\rhoi(\gammappp)$ since the canonical distribution is time-independent.
By combining all these results we finally get
\begin{eqnarray}
P_\mathrm{f}(p)&=&\int_{|p_x^{(\mathrm{f})}|\ge p}
d\bp^{(\mathrm{f})}d\br^{(\mathrm{f})}d\gamma^{(1)}
\rhom(\pppp)\,\rhop(\br)\,\rhoi(\gammappp)
\nonumber\\
&=&
\int_{|p_x^{(\mathrm{f})}|\ge p}
d\bp^{(\mathrm{f})}d\gamma^{(1)}\abs{\frac{p_x^{(\mathrm{f})}}{p_x^{(2)}}}\rhom(\pppp)\,\rhoi(\gammappp).
\label{e:Pp2}
\end{eqnarray}
To get the second line, we substituted (\ref{e:qtq}) and $\bp^{(0)}=\pppp$, and integrated over $\br^{(\mathrm{f})}$ (with $\bp^{(\mathrm{f})}$ and $\gamma^{(1)}$ fixed).
The factor $|{p_x^{(\mathrm{f})}}/{p_x^{(2)}}|$ is the Jacobian of the linear transformation  (\ref{e:qtq}). 
Surprisingly, we were able to carry out the $\br^{(\mathrm{f})}$ integral exactly since $\pppp$, $\rppp$ and $\gammappp$ do not depend on $\br^{(\mathrm{f})}$ and $\rhop(\br)$ is normalized.
The most delicate estimate is over.

We now use the energy conservation law
\begin{equation}
\frac{|\pppp|^2}{2m}+h(\gammappp)=\frac{|\bp^{(\mathrm{f})}|^2}{2m}+h(\gamma^{(1)})
\label{e:encon}
\end{equation}
and the explicit form of $\rhoi$ to get
\begin{eqnarray}
P_\mathrm{f}(p)&=&\int_{|p_x^{(\mathrm{f})}|\ge p}
\hspace{-10pt}
d\bp^{(\mathrm{f})}d\gamma^{(1)}\abs{\frac{p_x^{(\mathrm{f})}}{p_x^{(2)}}}\rhom(\pppp)
\frac{e^{-\beta h(\gamma^{(1)})-\frac{\beta}{2m}\{|\bp^{(\mathrm{f})}|^2-|\pppp|^2\}}}{Z(\beta)}
\nonumber\\
&\le&
\frac{3}{4\pi(\Delta p)^3}
\int_{|\bp^{(\mathrm{f})}|\ge p}
d\bp^{(\mathrm{f})}d\gamma^{(1)}\frac{|p_x^{(\mathrm{f})}|}{p_0-\Delta p}
\frac{e^{-\beta h(\gamma^{(1)})}}{Z(\beta)}
e^{-\frac{\beta}{2m}\{|\bp^{(\mathrm{f})}|^2-(p_0+\Delta p)^2\}},
\label{e:Pp3}
\end{eqnarray}
where we have enlarged the range of integral, and replaced $\pppp$ by its ``worst possible values'' and 
$\rhom(\pppp)$ by its maximum value $3/\{4\pi(\Delta p)^3\}$.
Now the $\bp^{(\mathrm{f})}$ integral and the $\gamma^{(1)}$ integral are completely
separated and we get
\begin{eqnarray}
P_\mathrm{f}(p)&\le&
\frac{3}{4\pi(\Delta p)^3}
\int_{|\bp^{(\mathrm{f})}|\ge p}
\hspace{-5pt}
d\bp^{(\mathrm{f})}\,
\frac{|\bp^{(\mathrm{f})}|}{p_0-\Delta p}\,
e^{-\frac{\beta}{2m}\{|\bp^{(\mathrm{f})}|^2-(p_0+\Delta p)^2\}}
\nonumber\\
&\le&
12\frac{(p+\pth)^3\pth}{p_0(\Delta p)^3}
e^{-\frac{\beta}{2m}\{p^2-(p_0+\Delta p)^2\}},
\label{e:Pp4}
\end{eqnarray}
where we recalled that $\pth=\sqrt{m/\beta}$, $p_0-\Delta p\ge p_0/2$, and bounded the
integral using the standard (and elementary) estimate
\begin{eqnarray}
\int_{|\bp'|\ge p}d\bp'\,|\bp'|\,e^{-|\bp'|^2/(2\alpha)}
&=&
\int_p^\infty dp'\,4\pi(p')^3\,e^{-(p')^2/(2\alpha)}
\nonumber\\
&=&\int_0^\infty ds\,4\pi(s+p)^{3}\,e^{-(s+p)^2/(2\alpha)}
\nonumber\\
&\le&
4\pi\,e^{-p^2/(2\alpha)}\int_0^\infty ds\,(s+p)^{3}\,e^{-s^2/(2\alpha)}
\nonumber\\
&\le&
8\pi\sqrt{\alpha}\,(p+\sqrt{\alpha})^{3}\,e^{-p^2/(2\alpha)},
\label{e:int}
\end{eqnarray}
where the final (crude) bound is checked by evaluating the integral explicitly.
By setting $\Delta p=\pth$ in (\ref{e:Pp4}), we get the desired (\ref{e:Pp0}).

To deal with the coefficient of restitution $\eR$, we note that for any $\delta>0$
\begin{equation}
\bkt{|p_x^{(\mathrm{f})}|}\le\bkt{|\bp^{(\mathrm{f})}|}\le p_0+\delta+
\int_{|\bp^{(\mathrm{f})}|\ge p_0+\delta}d\Gamma^{(\mathrm{f})}\,|\bp^{(\mathrm{f})}|\,\rho(\Gamma(\Gamma^{(\mathrm{f})})),
\label{e:pp}
\end{equation}
and bound the integral as before to get
\begin{equation}
\bkt{|\bp^{(\mathrm{f})}|}\le p_0+\delta+
9\sqrt{2\pi}\,\frac{\pth(p_0+\delta+\pth)^4}{(\Delta p)^3p_0}
e^{-\frac{\beta}{2m}\{(p_0+\delta)^2-(p_0+\Delta p)^2\}}.
\label{e:ppp2}
\end{equation}
By setting $\delta=2\Delta p=2\pth$, we get (\ref{e:R0}).

\section{Microcanonical distribution}
\label{s:MC}
We have proved our theorem in the setting where the internal degrees of freedom are initially distributed according to the canonical distribution.
To demonstrate that the choice of the equilibrium ensemble does not matter if the ball can be regarded as a ``healthy'' macroscopic system, we here show how to treat the case with the microcanonical distribution.
The following (not too short) analysis indeed corresponds to a single line calculation in (\ref{e:Pp3}).
From this one clearly sees that the canonical distribution is perfectly suited for a demonstration of the second law.

Let the density of states of the internal degrees of freedom be
\begin{equation}
\Phi(E):=\int d\gamma\,\delta(h(\gamma)-E).
\label{e:PhiE}
\end{equation}
We assume that $\Phi(E)$ satisfies the bounds
\begin{equation}
c_1\,\exp[N\,\sigma(E/N)]\le\Phi(E)\le c_2\,\exp[N\,\sigma(E/N)],
\label{e:PhiEexp}
\end{equation}
with a function $\sigma(\epsilon)$ which is smooth, nonincreasing, and concave in $\epsilon$, and constants $0<c_1\le c_2$.
Here $N$ is the number of particles in the ball, and is assumed to be a large quantity.
Of course $\sigma(\epsilon)$ is nothing but the entropy.
We note that the above assumption may not hold in general.
When this assumption is valid,  the system exhibits ``healthy'' thermodynamic behaviors, and the canonical distribution and the microcanonical distribution become equivalent in the sense that they give the same thermodynamics in the limit $N\to\infty$.
We shall now extend this equivalence to the case of the collision problem.

We then choose two constants $\epsilon_0>0$, $\DE>0$, and set the initial distribution of the internal degrees of freedom as
\begin{equation}
\rhoi(\gamma)=\frac{1}{\Omega_0}\,\chi[N\epsilon_0\le h(\gamma)\le
N\epsilon_0+\DE],
\label{e:riMC}
\end{equation}
where the characteristic function is defined by $\chi[\mbox{true}]=1$ and $\chi[\mbox{false}]=0$.
The normalization constant is
\begin{equation}
\Omega_0=\int_{N\epsilon_0}^{N\epsilon_0+\DE}dE\,\Phi(E)
\ge c_1\,\exp[N\,\sigma(\epsilon_0)]\,\DE.
\label{e:Omega0}
\end{equation}
This defines the microcanonical distribution with a finite width $\DE$.

Now we substitute the present $\rhoi(\gamma)$ to the right-hand side of \eqref{Pp2} to get
\begin{eqnarray}
P_\mathrm{f}(p)&=&\frac{1}{\Omega_0}\int_{|p_x^{(\mathrm{f})}|\ge p}
d\bp^{(\mathrm{f})}d\gamma^{(1)}\abs{\frac{p_x^{(\mathrm{f})}}{p_x^{(2)}}}\rhom(\pppp)\,
\chi[N\epsilon_0\le h(\gamma^{(2)})\le
N\epsilon_0+\DE]
\nonumber\\
&=&
\frac{1}{\Omega_0}\int_{|p_x^{(\mathrm{f})}|\ge p}
d\bp^{(\mathrm{f})}d\gamma^{(1)}\abs{\frac{p_x^{(\mathrm{f})}}{p_x^{(2)}}}\rhom(\pppp)\times
\nonumber\\
&&
\hspace{3cm}\times
\chi[N\epsilon_0\le h(\gamma^{(1)})+\frac{|\bp^{(\mathrm{f})}|^2}{2m}-\frac{|\bp^{(2)}|^2}{2m}\le
N\epsilon_0+\DE],
\label{e:PpMC1}
\end{eqnarray}
where we used the energy conservation \eqref{encon}.
We then note that for any $\bp^{(2)}=\bp^{(0)}$ such that $\rhom(\bp^{(2)})\ne0$, one has
\begin{equation}
\frac{(p_0)^2}{2m}-\DtE\le\frac{|\bp^{(2)}|^2}{2m}\le\frac{(p_0)^2}{2m}+\DtE
\label{e:DtE0}
\end{equation}
where
\begin{equation}
\DtE:=\frac{p_0\,\Delta p}{m}+\frac{(\Delta p)^2}{2m}.
\label{e:DtE}
\end{equation}
Then \eqref{PpMC1} and \eqref{DtE0} imply that 
\begin{eqnarray}
P_\mathrm{f}(p)&\le&
\frac{1}{\Omega_0}\int_{|p_x^{(\mathrm{f})}|\ge p}
d\bp^{(\mathrm{f})}d\gamma^{(1)}\abs{\frac{p_x^{(\mathrm{f})}}{p_x^{(2)}}}\rhom(\pppp)\times
\nonumber\\
&&
\hspace{1cm}\times
\chi[N\epsilon_0-\DtE\le h(\gamma^{(1)})+\frac{|\bp^{(\mathrm{f})}|^2}{2m}-\frac{|\bp^{(0)}|^2}{2m}\le
N\epsilon_0+\DE+\DtE]
\nonumber\\
&=&
\int_{|p_x^{(\mathrm{f})}|\ge p}
d\bp^{(\mathrm{f})}\abs{\frac{p_x^{(\mathrm{f})}}{p_x^{(2)}}}\rhom(\pppp)\times
\nonumber\\
&\times&
\frac{1}{\Omega_0}\int d\gamma^{(1)}
\chi[N\epsilon_0-\DtE\le h(\gamma^{(1)})+\frac{|\bp^{(\mathrm{f})}|^2}{2m}-\frac{|\bp^{(0)}|^2}{2m}\le
N\epsilon_0+\DE+\DtE].
\label{e:PpMC2}
\end{eqnarray}
Note that the $\bp^{(\mathrm{f})}$ integral and the $\gamma^{(1)}$ integral have been split.
We can evaluate the $\gamma^{(1)}$ integral separately as
\begin{eqnarray}
&&\frac{1}{\Omega_0}\int d\gamma^{(1)}\,
\chi[N\epsilon_0-\DtE\le h(\gamma^{(1)})+\frac{|\bp^{(\mathrm{f})}|^2}{2m}-\frac{|\bp^{(0)}|^2}{2m}\le
N\epsilon_0+\DE+\DtE]
\nonumber\\
&=&
\frac{1}{\Omega_0}\int dE\, \Phi(E)\,
\chi[N\epsilon_0-\DtE\le E+\frac{|\bp^{(\mathrm{f})}|^2}{2m}-\frac{|\bp^{(0)}|^2}{2m}\le
N\epsilon_0+\DE+\DtE]
\nonumber\\
&\le&
\frac{1}{\Omega_0}\,(\DE+2\DtE)\,
c_2\,\exp[N\,\sigma(\epsilon_0+\frac{1}{N}\{-\frac{|\bp^{(\mathrm{f})}|^2}{2m}+\frac{|\bp^{(0)}|^2}{2m}+\DE+\DtE\})]
\nonumber\\
&\le&
\rbk{1+\frac{2\DtE}{\DE}}\,
\frac{c_2\,\exp[N\,\sigma(\epsilon_0+N^{-1}\{-\frac{|\bp^{(\mathrm{f})}|^2}{2m}+\frac{|\bp^{(0)}|^2}{2m}+\DE+\DtE\})]}{c_1\,\exp[N\,\sigma(\epsilon_0)]}
\nonumber\\
&\le&
\rbk{1+\frac{2\DtE}{\DE}}\frac{c_2}{c_1}\,
\exp[\beta\{-\frac{|\bp^{(\mathrm{f})}|^2}{2m}+\frac{|\bp^{(0)}|^2}{2m}+\DE+\DtE\}],
\label{e:PpMC3}
\end{eqnarray}
where we used \eqref{Omega0} to get the second inequality.
To get the third inequality, we noted that \eqref{PhiEexp} and the concavity of $\sigma(\epsilon)$ imply
\begin{equation}
e^{N\,\sigma(\epsilon_0+(D/N))}
\le e^{N\,\sigma(\epsilon_0)+\beta D},
\label{e:Phiconv}
\end{equation}
for any $D$, where $\beta=\sigma'(\epsilon_0)$.

We now choose the uncertain $\DE$ (built into the microcanonical ensemble) comparable to $\DtE\simeq p_0\Delta p/m$.
Then by substituting \eqref{PpMC3} into \eqref{PpMC2}, we get essentially the same estimate as in the first line of (\ref{e:Pp3}) with an extra constant.
Note that the estimate gets worse as we take $\DE$ smaller.

\bigskip

It is a pleasure to thank Hisao Hayakawa, Tsuyoshi Mizuguchi, Hidetoshi Nishimori, Shin-ichi Sasa, and 
Ken Sekimoto for valuable discussions.
I also thank referees for their constructive comments on the manuscript.

%%%%%%%%%%%%%%%%%%%%%%%%%%%%%%%
\end{document}